\documentclass[showpacs,twocolumn]{revtex4}
\usepackage{graphicx}% Include figure files

\begin{document}

\title{Dynamic Scaling of Polymer Gels Comprising Nanoparticles}

\author{A. V. N. C. Teixeira}

\email{alvaro@ufv.br}

\affiliation{Universidade Federal de Vi\c cosa, Departamento de
F\'{\i}sica, CCE, 36570-900 Vi\c cosa, Brazil.}

\author{E. Geissler}

\affiliation{Laboratoire de Spectrom\'{e}trie Physique UMR CNRS
5588, Universit\'{e} Joseph Fourier de Grenoble, BP 87, 38402 St
Martin d'H\`{e}res cedex, France}

\author{P. Licinio}

\affiliation{Universidade Federal de Minas Gerais, Departamento de
F\'{\i}sica, ICEx, C.P.702, 30123-970 Belo Horizonte, Brazil.}

\date{\today}

\begin{abstract}
We present dynamic light scattering (DLS) measurements of soft
polymethyl-methacrylate (PMMA) and polyacrylamide(PA) polymer gels
prepared with trapped bodies (latex spheres or maghemite
nanoparticles). We show that the anomalous diffusivity of the
trapped particles can be analyzed in terms of a fractal Gaussian
network gel model for the entire time range probed by DLS technique.
This model is a generalization of the Rouse model for linear chains
extended for structures with power law network connectivity scaling,
which includes both percolating and uniform bulk gel limits. For a
dilute dispersion of strongly scattering particles trapped in a gel,
the scattered electric field correlation function at small
wavevector ideally probes self-diffusion of gel portions imprisoning
the particles. Our results show that the time-dependent diffusion
coefficients calculated from the correlation functions change from a
free diffusion regime at short times to an anomalous sub-diffusive
regime at long times (increasingly arrested displacement). The
characteristic time of transition between these regimes depends on
scattering vector as $\sim q^{-2}$ while the time decay power
exponent tends to the value expected for a bulk network at small
$q$. The diffusion curves for all scattering vectors and all samples
were scaled to a single master curve.
\end{abstract}

%%%\pacs{82.70.Gg, 83.10.Mj, 75.50.Mm}
% 75.50.Mm        - Magnetic liquids}
% 82.70.Gg Gels and sols
% 83.10.Mj Molecular dynamics, Brownian dynamics

\maketitle

\section{Introduction}

Dynamics of polymer and colloidal network systems has attracted
great interest in the last two decades \cite{adam:1988,
martin:1991, ren:1992, maity:1998, segre:1996, krall:1998,
cipelletti:2000}. Analysis of experimental results are almost
exclusively empirical, since theoretical analysis of these complex
systems is hindered by non-trivial intrinsic long range
correlations and non-ergodicity of the real networks. Complexity
arises from the visco-elastic and many-body structure of
interactions which extend over the whole system in gel networks
introducing memory effects. The topological constraints may
eventually lead to a localization of self-diffusion within a small
finite volume.

Dynamic light scattering (DLS) has served as a useful tool to probe
the molecular diffusion of this kind of system since the relaxation
of the density fluctuations is reflected in the dynamic scattered
electric field correlation function $g^{(1)}(t)$. Several works have
shown that on approaching the gel point the correlation function is
described as a simple exponential for short times and a stretched
exponential for longer times $\sim \exp \left[ -(t/\tau_c) ^\beta
\right]$, indicating an increasingly broad spectrum of relaxation
times \cite {adam:1988, martin:1991, ren:1992}. At the gel point the
system exhibits a critical slowing down while the correlation
function decays as a power law $\sim t^{-\alpha}$
\cite{martin:1991}. All regimes present power law dependence of the
parameters $\alpha$ and $\beta$ with the scattering vector $q$. One
of the interpretations of these results is based upon mechanisms of
anomalous diffusion, where the mean-square displacement $\langle
\Delta r^2(t) \rangle$ of scattering units is empirically given as
either a power law or a logarithmic regime \cite{ren:1992,
maity:1998}. The relation between the two functions is $g^{(1)}(t) =
\exp \left[ -q^2 \langle \Delta r^2(t) \rangle / 6 \right]$ where
$q$ is the scattering vector. In all analysis the temporal
correlation function of polymer systems is treated in separate parts
and a model that explains the dynamic behavior over the full time
range has not yet been presented. Similar results were observed in
colloidal gels \cite{krall:1998, cipelletti:2000}.

Beyond the gel point the correlation function tends, for large
times, to a non-zero plateau indicating a freezing of the density
fluctuations. The analysis of experimental correlation curves in
this case is given in terms of empirical expressions
\cite{martin:1991, maity:1998}; vibrational modes  of the network
\cite{krall:1998}; and also by the model of anomalous diffusion of
segments \cite {licinio:1997, maity:1998}. Krall and Weitz
\cite{krall:1998} also observed a plateau in the correlation
function for polystyrene colloidal gels. They analyzed the
correlation function in terms of overdamped normal modes of fractal
clusters where the mean square displacement is given by the sum of
exponential relaxations reflecting the relaxation of separate
localized elastic modes of the gel. This sum leads to a stretched
exponential type dependence:

\begin{equation}
\langle \Delta r^2(t) \rangle = \delta^ 2 [ 1 - e^{-(t/\tau)^p}].
\label{primeiraequacao}
\end{equation}
This expression describes the entire temporal domain seen by dynamic
light scattering. However the model does not fit the experimental
data well for long times in concentrated samples. It was also
observed that the parameters $\delta^2$, $\tau$ and $p$ depend on
both the scattering angle and the gel concentration.

In this letter we use dynamic light scattering to measure the
internal dynamics of chemically cross-linked polymer gels. We use a
Gaussian network model that differs from the Krall-Weitz model in
considering the relaxation scaling. This model allows for a good fit
of the experimental curves in all the observable time range while
giving an interpretation for the fitting parameters.

\section{Experimental section}

\subsection{Materials and methods}

We prepared a sample of aqueous poly(methyl methacrylate) gels by
co-polymerization of methyl methacrylate (MMA, Aldrich) with
ethylene glycol dimethacrylate (EGDM, Aldrich). The ratio MMA/EGDM
was fixed at 50/1. To initiate the polymerization reaction, a small
quantity of $\alpha-\alpha'$ azobisisobutyronitrile was added. The
gels were prepared with 10\% wt. of solute (MMA + EGDM). Before
adding the initiator, a small portion of standard latex spheres
(Duke Scientific)  of diameter 0.198 $\mu$m was incorporated.  The
final solution was homogenized by vigorous shaking. We will refer to
this sample as Latex.

We also prepared three ferrogel samples, also known as magnetic gels
or magnetic elastomers, with aqueous poly(acrylamide) and industrial
ferrofluids. The gels were made by co-polymerization of acrylamide
with $N$,$N$' methylene bisacrylamide (Acros) in the presence of
ammonium persulfate and $N$,$N$,$N$',$N$'-tetramethylenediamine
\cite{morris}. The ratio acrylamide / bisacrylamide was fixed at
30/1 and the total monomer concentration (acrylamide +
bisacrylamide) was fixed at 2.5\% wt. To these gels a small portion
of aqueous magnetite ferrofluid was added. In the first sample,
defined as M300, we used a M300 ferrofluid obtained from Sigma
Hi-Chemical in such a way as to have a final concentration of $1.7
\times 10^{-3}$\% wt. of solids. The second and third samples,
defined as EMG408 and EMG308, correspond to gels with the
ferrofluids EMG408 and EMG308, both obtained from Ferrotec, having a
final solid concentration of $3 \times 10^{-3}$\% wt. and $2 \times
10^{-3}$\% wt., respectively. The M300 ferrofluid was stabilized
with the surfactant SDSB (sodium dodecyl-benzene sulfate). The type
of surfactant employed in the Ferrotec ferrofluid was not disclosed
by the supplier.

All samples were prepared in sealed 20 mL cylindrical glass vials
(Wheaton) and were kept at $\sim 80^\circ$C for nearly 12 hours to
ensure that all monomers had reacted.

Dynamic light scattering measurements were made using a 20 mW He-Ne
laser ($\lambda = 6328$ \AA) and a photomultiplier as a detector.
The detector was placed in a goniometer and the angle of scattering
was controlled by a computer. The correlation functions were
calculated by a BI9000-AT correlator board by Brookhaven Inst. Co.
with 522 channels. During the measurements the samples were
maintained in a thermal bath at 25.0$^\circ$C with a precision of
$\sim$ 0.1$^\circ$C.

To minimize non-ergodic sampling effects inherent in polymer gels,
we adopted the strategy of sampling a large gel volume and used a
relatively large detection area in the photomultiplier (pin-hole
diameter 400 $\mu$m, corresponding to nearly 15 coherence areas).
Signal to noise ratios close to $10^3$ could be obtained in
measurements of 1 to 3 hours each.

\subsection{Data analysis}

DLS experiments yield the normalized intensity correlation function
$g^{(2)}(t) = \langle I(0).I(t) \rangle / \langle I \rangle^2$,
which is related to the electric field correlation function
$g^{(1)}(t) = \langle E(0).E^*(t) \rangle / \langle I \rangle$. For
stationary fluctuations with a Gaussian distribution, in the
so-called homodyne mode, the two correlation functions are related
by \cite{berne}

\begin{equation}
g^{(2)}(t) = 1 + \beta \vert g^{(1)}(t) \vert^2 \label{sierge}
\end{equation}
where $\beta$ is an experimental constant given essentially by the
number of coherence areas.

In the heterodyne mode, only part of the total intensity
fluctuates, while an independent static coherent source (local
oscillator), is also considered. In this case one has

\begin{equation}
g^{(2)}(t)=1 + \beta \left[ \frac{\langle I_{f}
\rangle^{2}}{\langle I \rangle^2} \vert g^{(1)}(t) \vert^{2} +
2\frac{\langle I_{f} \rangle \langle I_{s} \rangle}{\langle I
\rangle^2} \vert g^{(1)}(t) \vert \right]. \label{heter}
\end{equation}
Here $\langle I_f \rangle$ and $\langle I_s \rangle$ are the mean
value of the fluctuating and static part of the scattered light,
respectively, and $\langle I \rangle = \langle I_f \rangle +
\langle I_s \rangle$ is the total intensity.

In polymers and gels, the self-diffusivity of molecular chains is
progressively slowed, since to perform any movement they have to
drag surrounding polymer segments. This effect is more pronounced as
the number of connections or complexity of the structure increases.
For highly connected networks the diffusivity may reach a state of
bounded diffusion, in which each segment diffuses only in a limited
region of space. In previous works we have shown that the diffusion
regime is determined by the topological connectivity in the network
given by its spectral dimension $d_s$. \cite{avteixeira:1999,
licinio:2001}. Accordingly, a critical value $d_s = 2$ divides the
non-bounded from the bounded self-diffusion regime. The consequence
of a bounded regime in DLS experiments is that the localization of
gel segment excursions leads to a mean frozen structure. This also
corresponds to the non-ergodic nature of gels leading to persistent
speckle patterns in the scattered intensity. Above the gel point, as
the structure progressively freezes, it becomes less correlated, the
fluctuations being localized within independent volumes given by
mesh sizes typically less than 1 micron. Dynamics of density or
structure fluctuations around the mean can be understood as the
source of a dynamic signal superimposed on the mean speckles. The
scattering becomes intrinsically heterodyne.

To obtain the information about the dynamics of the polymer gels one
considers both the fluctuating and static parts in the normalized
first order correlation function. This is achieved by defining the
gel correlation function as

\begin{equation}
g^{(1)}_{gel}(t) = \frac{\langle I_s \rangle + \langle I_f \rangle
\; g^{(1)}(t)}{\langle I \rangle} = (1-Y) + Yg^{(1)}(t),
\label{g1gel}
\end{equation}
while $g^{(1)}(t)$  contains only the fluctuating part of the
scattered field as in Eq.\ref{heter}. $Y$ measures the intensity
fluctuation strength and is defined as $Y \equiv \langle I_f \rangle
/\langle I \rangle $. The definition above implicitly decouples the
static field from the fluctuating field at the detector. Hence the
static field is defined as the average of the scattered electric
field complex amplitude, which corresponds to the mean position of
the confined scatterers.

Replacing $g^{(1)}(t)$ of Eq.\ref{g1gel} in Eq.\ref{heter} one
obtains the relation between the intensity (measured) and electric
field correlation function:

\begin{equation}
g^{(2)}(t)= 1 + \beta\vert g^{(1)}_{gel}(t) \vert^2 - \beta
(1-Y)^2. \label{siergemod}
\end{equation}

The electric field correlation functions $g^{(1)}_{gel}(t)$ are
calculated from Eq.\ref{siergemod}, in which the parameter $\beta$
was found for each scattering angle in the limit $\beta = \lim_{t
\rightarrow 0}[g^{(2)}(t) - 1]$ measured in a dilute solution of the
same standard latex used in the gel, whose value changes slightly
with the detector angle and lies around 0.5. The small value of
$\beta$ is due to the large detection area. The fluctuation strength
$Y$ was obtained for each scattering angle by extrapolation to $t
\rightarrow 0$ in Eq.\ref{siergemod}, noting that $g^{(1)}_{gel}(0)
= 1$. So $Y = \lim_{t \rightarrow 0} 1 - \left[ 1- \left(
g^{(2)}(t)-1 \right) / \beta \right]^{1/2}$.

\section{Results and discussion}

\begin{figure}[hbt]
\includegraphics[scale=0.35, angle=0]{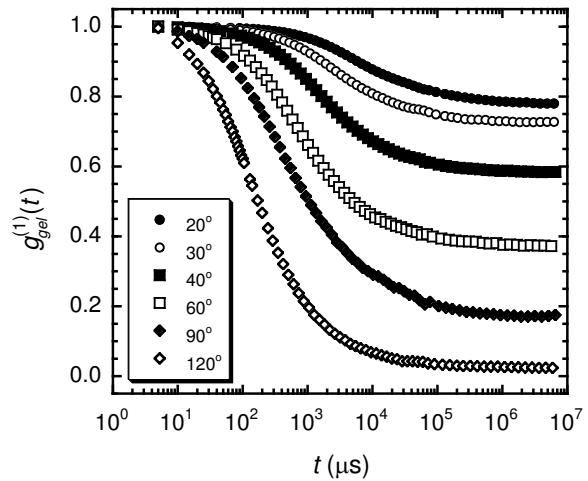}
\caption{Electric field correlation function for methyl
methacrylate gels with latex spheres} \label{correl}
\end{figure}

A set of typical correlation functions $g^{(1)}_{gel}(t)$ are shown
in Fig.\ref{correl}. In all samples the correlation curves show an
exponential decay at short times ($t \stackrel{<}{_\sim} 100 \mu s$)
and an asymptotic behavior at very long times. As discussed by Krall
and Weitz \cite{krall:1998} this behavior can be explained as the
result of the restricted movement of scattering particles. Since the
particles are allowed to visit only a limited volume, their
positions as well as the scattered intensity in different times are
always correlated. The asymptotic correlation value depends on the
ratio of the length through which the particles can diffuse to the
scattering length $q^{-1}$. The scatterers must change their
relative position by about $q^{-1}$ to make an appreciable variation
in the scattered intensity. So in the condition of large $q^{-1}$
(small angles) the restricted particle movements have little effect
in changing the value of the scattered intensity and and small
fluctuations result. In the opposite small $q^{-1}$ limit, the
intensity fluctuations become quasi homodyne and the correlation
function decays fully. Since the system is non-ergodic, we
calculated the average from approximatively 20 different positions
of one of the samples to obtain realistic ensemble averages. However
we verify that observation of different positions of the same sample
changes only the asymptotic plateau, but not the shape of the
correlation curves. In all the other samples the data was collected
in just one position.

Since the light scattered by the particles-containing gels is at
least 20 times greater than that from neat gels, it can be assumed
that the signal is due essentially to the trapped particles. In
dilute conditions the correlation function is related to the mean
square displacement of each scattering particle $\langle \Delta
r^2(t) \rangle$ by $g^{(1)}_{gel}(t)=\exp \left( -q^2 \langle \Delta
r^2(t) \rangle / 6 \right)$ and we can calculate the self-diffusion
coefficient from the correlation function by

\begin{equation}
D(t) = \frac{1}{6} \frac{\partial \langle \Delta r^2(t)
\rangle}{\partial t} = - \frac{1}{q^2}\frac{\partial \ln
g^{(1)}_{gel}(t)}{\partial t} \label{ddef}
\end{equation}

\begin{figure}[hbt]
\includegraphics[scale=0.35, angle=0]{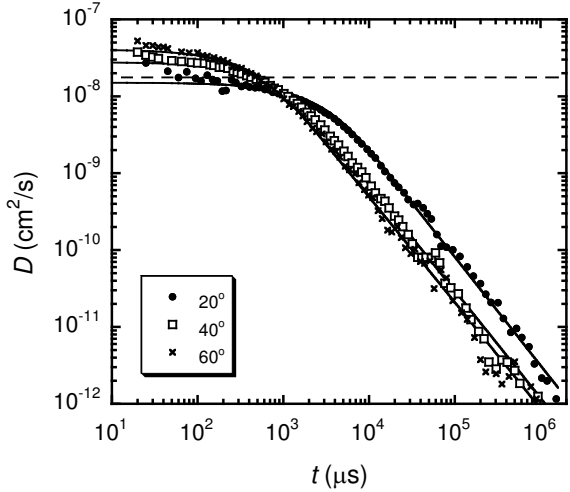}
\caption{Self-diffusion coefficient for latex probes in methyl
methacrylate gels. The continuous lines are the least square fits to
Eq.\ref{drouse}. The dashed line is the expected value for the free
diffusion of the probes.} \label{dif}
\end{figure}

In Fig.\ref{dif} it is shown that the diffusion coefficient $D(t)$
has two distinct regimes: free diffusivity for short times where
$D(t)$ is nearly constant; and an anomalous regime characterized by
a temporal power law over almost three decades.

To describe the full-time behavior of $D(t)$ we use a mass-spring
Rouse-like or Gaussian network model \cite{doi,grosberg}. In
previous works we generalized the Rouse model to describe the
dynamics either of regular networks with arbitrary dimensionality
\cite{licinio:1997, licinio:1998} or that of fractal networks
\cite{avteixeira:1999, avteixeira:tese}. For regular networks, it was shown
that the mean square displacement for a network of
dimensionality $d$ with $N^d$ bond units is given by

\begin{eqnarray}
\left\langle \Delta r^2(t) \right\rangle &=& \frac{6 D_0}{N^d} t +
\frac{6 D_0}{N^d} \tau N^2 \Bigg[ \sum_{p_1=1}^N \frac{1 - e^{- t
p_1^2
/\tau N^2}}{p_1^2} \; + \nonumber \\
&+& \sum_{p_1, p_2=1}^N       \frac{1 - e^{-t (p_1^2 +p_2^2)/\tau
N^2}}
{p_1^2 + p_2^2} \; \ldots \nonumber \\
&+& \sum_{p_1,\ldots,p_d=1}^N \frac{1 - e^{-t (p_1^2 + \ldots +
p_d^2)/\tau N^2}}{p_1^2 + \ldots + p_d^2} \Bigg], \label{r2}
\end{eqnarray}
where $\tau$ is the single elastic-link relaxation time and $D_0$
is the free bond-particle diffusion constant.

The expression for the mean square displacement predicts three
distinct regimes: for short times ($t \ll \tau$) the last term tends
to a linear dependence on time, which corresponds to the free
diffusion regime of the isolated units with $\left\langle \Delta
r^2(t) \right\rangle \approx 6D_0t$; for long times the first term
is dominant and the mean square displacement is given by the
diffusion of the entire structure with mass proportional to $N^d$ or
$\left\langle \Delta r^2(t) \right\rangle \approx 6D_0t/N^d$. This
regime is not achieved experimentally for ordinary gels, but it
could be verified for isolated clusters or microgels.

From Eq.\ref{r2} it is straightforward to show that the diffusion
coefficient in Eq.\ref{ddef} is then given by

\begin{equation}
D(t) = \frac{D_0}{N^{d}} \left[ 1 + x(t) + x(t)^2 + \ldots +
x(t)^{d} \right] \label{difftotal}
\end{equation}
where $x(t)$ is the sum of the relaxations of the $N$ vibration
modes

\begin{equation}
x(t) = \sum_{p=1}^N e^{-tp^2/\tau N^2} \label{sdef}.
\end{equation}

Evaluating the terms in Eq.\ref{r2} and Eq.\ref{difftotal} we
verified that the last sum is dominant for infinitely large networks
($N \rightarrow \infty$), and hence

\begin{equation}
D(t) \approx D_0 \left[ \frac {x(t)}{N} \right] ^{d}
\label{drouse}.
\end{equation}

As for the mean square displacement, at short times ($t \ll \tau$)
the diffusion coefficient approaches the free diffusion
coefficient $D_0$. For very long times the sum in Eq.\ref{sdef}
can be replaced by an integral and, for large networks
\begin{equation}
x(t) \approx \int_0^\infty e^{-tp^2/\tau N^2} \; dp = N \left(
\frac {4t}{\pi \tau} \right)^{-1/2}
\end{equation}
which leads to the anomalous regime with
\begin{equation}
D(t) = D_0 \left( \frac {4t}{\pi \tau} \right)^{-d/2}.
\label{danomalous}
\end{equation}
corresponding to
\begin{equation}
\left\langle \Delta r^2(t) \right\rangle = \left\langle \Delta
r^2_\infty \right\rangle - B t^{1-d/2} \label{r2anomalous}
\end{equation}
where $\left\langle \Delta r^2_\infty \right\rangle$ is the
asymptotic value of the mean square displacement and $B$ is a
constant.

The same equations are valid for fractal networks, replacing $d$ in
Eq.\ref{drouse}, Eq.\ref{danomalous} and Eq.\ref{r2anomalous} with
the spectral dimension $d_s$, which describes the network topology.
The data presentation using $D(t)$ instead of $\langle \Delta r^2(t)
\rangle$ is preferable since the anomalous regime is easily
recognized as a straight line in a log-log plot. In Fig.\ref{dif} we
can see the least squares fit to the experimental data using
Eq.\ref{drouse} (continuous lines), with $N=1000$, which is a very
good approximation to the limit $N \rightarrow \infty$. The model
gives a fair description of the diffusion dynamics over the whole
time range of the measurements. In the same plot is shown the
expected value for the free diffusion coefficient for the latex
beads in pure solvent (dashed line). We observe that the diffusion
coefficient of the gels in the thermodynamic limit $q \rightarrow 0$
using latex beads corresponds approximately to the expected value
for the free diffusion coefficient. In the ferrogels the measured
limiting diffusion coefficient is approximately one order of
magnitude smaller than the free diffusion value. It was observed
that some ferrofluids form aggregates when dispersed in a gel matrix
\cite {avteixeira:2003pre} thereby decreasing the apparent diffusion
coefficient. However the size of the aggregates is not large enough
to explain the measured limiting values ($D_0 \sim 4\times 10^{-9}$
cm$^2$/s for the EMG308; $D_0 \sim 1\times 10^{-8}$ cm$^2$/s for the
EMG408 and $D_0 \sim 2\times 10^{-9}$ cm$^2$/s for the M300). It is
still plausible to account for network distortions around the
magnetic particle aggregates that increase their effective viscous
drag and decrease their effective initial diffusivity.

\begin{figure}[hbt]
\includegraphics[scale=0.35, angle=0]{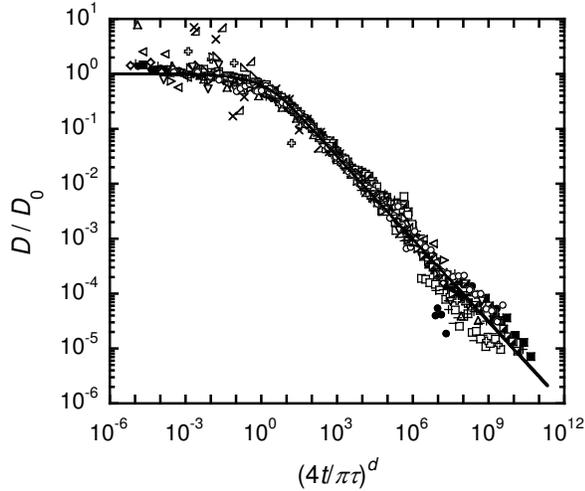}
\caption{Scaled self-diffusion coefficient for all samples at all
scattering angles.} \label{dscale}
\end{figure}

\begin{figure}[hbt]
\includegraphics[scale=0.35, angle=0]{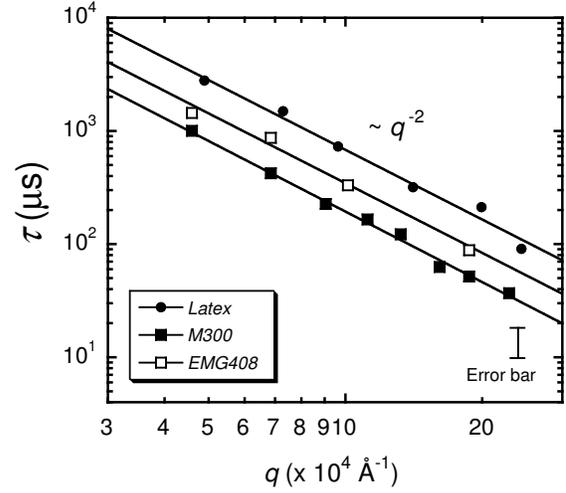}
\includegraphics[scale=0.35, angle=0]{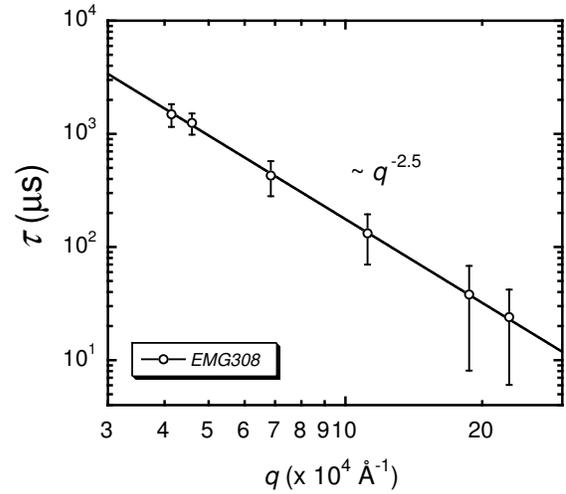}
\caption{The parameter $\tau$ as function of $q$. The continuous
lines are power law regimes $\tau \sim q^{-y}$ with $y=2$ for the
samples: Latex, M300 and EMG408 and $2.5$ for the EMG308.}
\label{tau}
\end{figure}

The curves for all the samples can be scaled into a single master
curve (Fig.\ref{dscale}) consistent with the predicted
behavior of fractal Rouse networks.
Although the form of the diffusion coefficient remains the same for
all scattering angles, the fitted parameters display a $q$ dependence.
The relaxation time, reflecting the network coarsening at different
scattering wavelengths, follows a power law $\tau \sim q^{-y}$ with
$y$ close to 2 in three samples and $y=2.5\pm 0.3$ in one
sample, as shown in Fig.\ref{tau}. From Eq.\ref{ddef} it is easy to
show that the anomalous regime (Eq.\ref{danomalous}) corresponds to
a stretched exponential behavior normally observed in gel systems,
which is given by
\begin{equation}
g^{(1)}_{gel}(t) \propto \exp \left[ -\left(
\frac{t}{\tau^*}\right)^{1-d/2} \right]
\end{equation}
with $\tau^* \sim  (q^4\tau^d)^{1/(d-2)}$. In the case of $\tau \sim
q^{-2}$ the characteristic time $\tau^*$ has has the same power
regime $\tau^* \sim q^{-2}$ for all $d$. It has been found in other
works that the power exponent for $\tau^*$ varies between 2 and 3,
taking the values $3.0\pm 0.4$ and $2.5\pm 0.2$ for gelatin,
depending on the sample temperature \cite{ren:1992}; $2.05 \pm 0.08$
for gelatin above the gel point and $3.0 \pm 0.2$ beyond the gel
point \cite{maity:1998}; while for silica gels it was found to be $y
\approx 2$ \cite{martin:1991}.

One can argue that the transition time $\tau$ from free diffusion to the
anomalous diffusion depends on the length scale defined by
the inverse of the scattering vector. For a viscoelastic medium
of size $q^{-1}$ the relaxation time is given by the ratio of the
friction factor and the elastic constant $\zeta_q / \kappa_q$. For
regular 3-D networks in the free draining limit we expect the
friction factor to be proportional to the number of units, i.e., to the
volume, and the elastic constant to be linear in the length scale.
Thus
\begin{eqnarray}
\zeta_q \sim q^{-3} \nonumber \\
\kappa_q \sim q^{-1}
\end{eqnarray}
which leads to $\tau_q \sim q^{-2}$, corresponding to results
found in our experiments.

\begin{figure}[hbt]
\includegraphics[scale=0.35, angle=0]{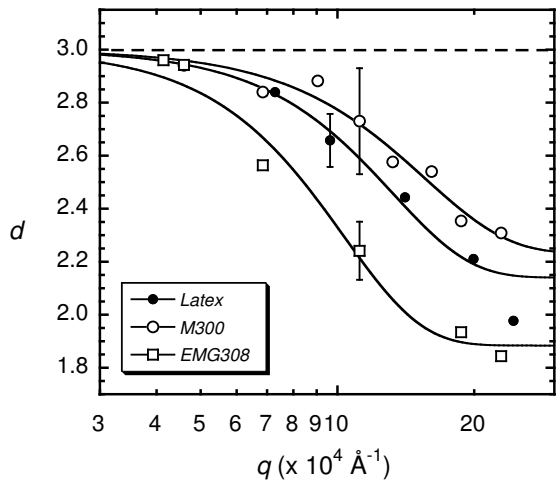}
\caption{The parameter $d_s$ as function of $q$. The continuous line
is a guide for the eye and the dashed line shows the expected value
for regular networks. Typical error bars are shown.} \label{ds}
\end{figure}

It can be seen, however, that these networks are far removed from the
regular homogeneous model. Gel structure is reflected in the power
exponent of the diffusion coefficient in the anomalous regime, or
the spectral dimension. It is shown in Fig.\ref{ds} that the
spectral dimension for three samples lies between 1.8 and 3 and at
small wavenumbers it tends to the value expected for regular
3-D networks, $d_s = 3$. This apparent discrepancy indicates some
cancelation in fractal scaling corrections to visco-elastic
effects, giving a weak fractal dimension dependence on the
relaxation-size scaling of the networks, thus remaining the
typical 3-D $q^{-2}$ dependence.

\section{Conclusions}

In this paper we have reported dynamic light scattering measurements
of polymer gels in which monodisperse 0.2 $\mu$m latex particles and
ferromagnetic particles are used as probes. In this way the
correlation function gives information about the probe dynamics.
Since the particles are confined by the surrounding network, their
movements also reflect the local dynamics of gel segments.
Experimental correlation function analysis is performed by
normalizing the electric field correlation function to include the
intrinsic static contribution of the gel. The temporal behavior can
be better understood by displaying the time-dependent diffusion
coefficient. Two regimes are observed - free diffusion and anomalous
diffusion - for all samples at all scattering angles. These regimes
are modeled in a consistent way by the simple mass-spring or
Gaussian network model over the entire time range. Analyzing the
characteristic time that separates the two regimes as a function of
scattering wavenumber, we find a power law regime for most of the
samples with an exponent close to -2. Similar dependence has already
been reported for different systems exhibiting long-range
organization. Our results also show that the dynamics of the gels
approaches the behavior expected for regular 3-D networks at large
length scales.

\section{Acknowledgements}

This work was supported by the Brazilian agencies CNPq, FAPEMIG
and CAPES.

\end{document}